\documentclass[prb,twocolumn,showpacs,preprintnumbers,amsmath,amssymb]{revtex4}
\usepackage{graphicx}
\usepackage{amsmath}
\usepackage{amsbsy}
\usepackage{dcolumn}
\usepackage{bm}

\begin{document}

\newcommand{\w}{\omega}
\newcommand{\sign}{\text{sign}}
\newcommand{\s}{\sigma}
\newcommand{\RE}{\text{Re}}
\newcommand{\IM}{\text{Im}}
\newcommand{\TK}{T_K}
\newcommand{\g}{\gamma}
\newcommand{\G}{\Gamma}
\newcommand{\Nf}{N_F}
\newcommand{\pf}{PF}
\renewcommand{\vec}[1]{{\mathbf #1}}


\title{
Spectral function of the Kondo model in high magnetic fields}

\author{A. Rosch, T. A. Costi, J. Paaske,  and P. W\"olfle}

\affiliation{Institut f\"ur Theorie 
der Kondensierten Materie, Universit\"at Karlsruhe, 
D-76128 Karlsruhe.}

\date{\today}

\begin{abstract}
  Using a recently developed perturbative renormalization group (RG)
  scheme, we calculate ana\-ly\-ti\-cal\-ly the spectral function of a
  Kondo impurity for either large frequencies $\w$ or large magnetic
  field $B$ and arbitrary frequencies.  For large $\w\gg \max[B,T_K]$
  the spectral function decays as $1/\ln^2[\w/T_K]$ with prefactors
  which depend on the magnetization.  The spin-resolved spectral
  function displays a pronounced peak at $\w \sim B$ with a
  characteristic asymmetry.  In a detailed comparison with results
  from numerical renormalization group (NRG) and bare perturbation
  theory in next-to-leading logarithmic order, we show that our
  perturbative RG scheme is controlled by the small parameter
  $1/\ln[\max(\w,B)/T_K]$.  Furthermore, we assess the ability of the
  NRG to resolve structures at finite frequencies.
\end{abstract}

                                 

\maketitle

\section{Introduction}
Only recently it has become possible to measure
directly\cite{STM,Goldhaber98,Franceschi} the so-called Kondo
resonance, i.e. a sharp resonance characteristic for the spectral
function of a single magnetic impurity in a metal.  
The Kondo resonance is probably the most direct mani\-fes\-tation of the
Kondo effect\cite{Hewson93}: The antiferromagnetic coupling between
conduction electrons and a localized spin causes a complex
many-body resonance which ultimately  screens the magnetic degree of
freedom.
The spectral function is directly proportional to a tunneling
current, e.g. from the tip of a scanning tunneling microscope to a
magnetic impurity on the surface of a metallic host\cite{STM}.
Alternatively it can be measured by  tunneling through weak links
into a quantum dot in the Kondo regime\cite{Goldhaber98,Franceschi}.
This also opens the possibility of investigating the spectral function
out of equilibrium\cite{Franceschi}.

While the spectral function is an interesting quantity in its own right, it
also has many important applications. For example, it is directly
related to the $T$-matrix, which describes the scattering of the
conduction electrons (see below). It therefore determines not only the
transport properties of magnetic impurities and quantum dots in the
Kondo regime but it also plays an important role in controlling the
distribution\cite{dephasing} of electrons in metals contaminated by
magnetic impurities.  Furthermore, the spectral function of impurity
models is central in determining the properties of lattice models within
dynamical mean field theory\cite{DMFT}.

The spectral function of the Kondo model in a magnetic field has been
calculated using a variety of
techniques
\cite{Hewson93,Meir,Sakai,Takagi,costi.00,hofstetter.00,MooreWen,%
  Logan,Hewson01,Martinek}, including diagrammatics, mean-field treatments and
quantum Monte-Carlo.  The most precise method, however, is the
numerical renormalization group\cite{wilson.75+kww.80} (NRG) and its
generalizations\cite{Frota,sakai.89,costi.94,hofstetter.00} which
allow the calculation of dynamical quantities. However, little is
known analytically about the properties of the spectral function,
particularly for high magnetic fields. For small $B$, the spectral
function for small frequencies can be calculated from Fermi liquid
theory and renormalized perturbation theory\cite{Logan,Hewson01}. We
have recently developed\cite{RoschRG} a perturbative renormalization
group scheme based on frequency-dependent couplings for the Kondo
model.  While this method was formulated to describe the Kondo effect
out of equilibrium, it can also be used to calculate dynamical
quantities in equilibrium in a controlled way for either large
frequencies or large magnetic fields.  We therefore use this method to
determine analytically the spectral function in leading order of
$1/\ln[\max(B,\w)/T_K]$, focusing our attention on equilibrium and
vanishing temperature. This clarifies some long-standing questions on
the asymptotics of the spectral function in the Kondo model. A
detailed comparison with results from NRG and with perturbation theory
allows us to check the assumptions underlying our perturbative RG
approach.

\section{The model}

We consider the spectral function $\IM G^f_\sigma(\w)$ of an electron in an impurity
orbital  within the Anderson model
\begin{multline}
 H_A=\sum_{{\bf k},\s} \varepsilon_{{\bf k}}
c^{\dagger}_{{\bf k}\s}c_{{\bf k}\s}+\sum_{\s}\epsilon_f 
f^\dagger_{\sigma} f_{\sigma}
\\
+
V \sum_{\vec{k},\sigma} (c^\dagger_{\vec{k}\sigma} f_\sigma+h.c.)+U n_{f,\uparrow}
 n_{f,\downarrow}
\end{multline}
where $c^\dagger_{\vec{k}\sigma}$ and $f^\dagger_\sigma$ are creation
operators for electrons in the conduction band and in the impurity
orbital, respectively, $n_{f\sigma}=f^\dagger_\sigma f_\sigma$, and
$U$ is a large Coulomb interaction matrix element.  We focus our
attention on the Kondo regime, $\epsilon_f \ll -V^2 N_F$,
$\epsilon_f+U \gg V^2 N_F$ ($N_F$ being the conduction electron
density of states), where the impurity orbital is occupied by a
localized spin $\vec{S}$ (with $S=1/2$) and the problem can be mapped
onto the Kondo Hamiltonian
\begin{multline}
\label{kondo}
H_K=\sum_{{\bf k},\s} \varepsilon_{{\bf k}} c^{\dagger}_{{\bf
    k}\s}c_{{\bf k}\s}- B S_{z} + J \!\!\! \sum_{{\bf k},{\bf
    k}',\s,\s'} \!\!\!\! \vec{S} \cdot c^{\dagger}_{{\bf k}'\s'}
{\boldsymbol \tau}_{\s'\s}c_{{\bf k}\s}.
\end{multline} 
where ${\boldsymbol \tau}$ is the vector of Pauli matrices. Note, that we have coupled an extra
magnetic field $B$ to the impurity\cite{magneticField}, measuring $B$
in units of the Zeeman splitting.

The Green function $G^f_\sigma$ within the Anderson model is directly
related to the $T$-matrix $T_\sigma(\w)=V^2 G^f_\sigma(\w)$ of the conduction
electrons, defined by
\begin{eqnarray}
g^c_{\sigma,\vec{k},\vec{k}'}(\w)&=&g^0_{\vec{k}}(\w) \delta(\vec{k}-\vec{k}')+
g^0_{\vec{k}}(\w) T_\sigma(\w) g^0_{\vec{k}'}(\w)
\end{eqnarray}
where $g^c_{\sigma,\vec{k},\vec{k}'}$ ($g^0_{\vec{k}}$) is the (bare)
Green function of the conduction electrons.  While it is not
completely obvious how to define a spectral function for a Kondo
impurity, the $T$-matrix of the Kondo model can easily be identified
with the following correlation function, e.g. 
using equations of motion\cite{costi.00},
\begin{multline}
T_\sigma(\w)= V^2 G^f_\sigma(\w)=J \langle S_z \rangle +J^2
\left\langle\!\left\langle
 \vec{S} c^\dagger_\alpha {\boldsymbol \tau}_{\alpha \s};
  \vec{S}  {\boldsymbol \tau}_{\s \alpha'} c_{\alpha'} 
\right\rangle\!\right\rangle, \\ \label{eom}
\end{multline}
where $\langle\langle...\rangle\rangle$ denotes a retarded correlation
function.  Formally, we will consider the usual
``scaling limit'', where the ratios of frequency or magnetic field and
Kondo temperature, $\w/T_K, B/T_K$, are kept
fixed while the ratio of $T_K$ and all ``high energy'' scales like
bandwidth $D_0$, level position or interaction are sent to zero,
$T_K/D_0, T_K/U,T_K/\epsilon_f\to 0$. In this limit, the dimensionless
product of the $T$-matrix and the density of states, $N_F \IM T(\w)$, is a
universal function of $\w/T_K$ and $B/T_K$ and it coincides with the
low-frequency part of the spectral function $N_F V^2 \IM
G^f_\sigma(\w)$ of the Anderson impurity.  Therefore $N_F \IM
T_\sigma(\w)$ is the quantity to be studied below.

\section{Perturbative RG}

The perturbation theory of the Kondo model is characterized by
logarithmic divergences. The method of choice for the resummation of
leading logarithmic corrections is the perturbative renormalization
group (RG). In Ref.~\onlinecite{RoschRG} we have developed a certain
formulation of perturbative RG based on frequency-dependent coupling
constants. A thorough discussion of this approach will be published
elsewhere\cite{Unpubl}; here we will use the RG equations without
further derivation and check the results extensively in a comparison
with bare perturbation theory and NRG.

\begin{figure}
$\displaystyle \frac{\partial}{\partial \ln D}
$\begin{minipage}{0.7cm}\includegraphics[height=0.7cm]{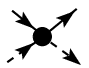}
\end{minipage}
\begin{minipage}{6cm}
\includegraphics[width=\linewidth]{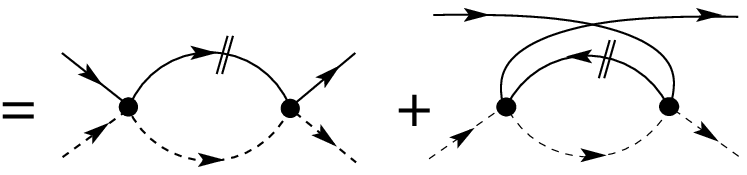}
\end{minipage}
\caption{\label{diagram}
Diagrammatic form of the RG equation (\ref{jRG}). The strokes symbolize derivatives with respect to $\ln D$, see 
Ref.~\onlinecite{RoschRG} for details.}
\end{figure}
The RG is formulated in terms of frequency- and spin-dependent
vertices\cite{pseudo} $g^{\sigma\w_c;\sigma'\w_c'}_{\gamma\w_f;
  \gamma'\w_f'}$ where $\sigma$ and $\w_c$ ($\gamma$ and $\w_f$)
denote the spin and energy of the incoming electron (the incoming
spin), see Ref.~\onlinecite{RoschRG}.  Primed quantities refer to
outgoing particles. Before renormalization starts, the vertex
$g^{\sigma\w_c;\sigma'\w_c'}_{\gamma\w_f; \gamma'\w_f'}$ is just given
by the bare coupling $N_F J {\boldsymbol \tau}_{\gamma \gamma'}
{\boldsymbol \tau}_{\sigma \sigma'}$ independent of frequencies.  In
leading order of the relevant expansion parameter $1/\ln[
\max(\w,B)/T_K]$, one can set the energy of the spin on-shell
$\w_f=-\gamma B/2, \w_f'=-\gamma' B/2$ to keep track only of the
energy of the incoming electron. The vertex is parametrized as
\begin{multline}
g^{\sigma,\w;\sigma',\w-(\gamma-\gamma') B/2}_{\gamma,-\gamma B/2;
\gamma',-\gamma' B/2}=\tau^z_{\gamma \gamma'} \tau^z_{\sigma \sigma'}
\tilde{g}_{z\sigma}(\w) \label{vertex} \\
+
(\tau^x_{\gamma \gamma'} \tau^x_{\sigma \sigma'}+
\tau^y_{\gamma \gamma'} \tau^y_{\sigma \sigma'}) \, 
\tilde{g}_{\perp}(\w- \gamma B/2).
\end{multline}
The one-loop RG equations (see Fig.~\ref{diagram}) read\cite{RoschRG} in terms of the running cut-off $D$
\begin{eqnarray}
\frac{\partial \tilde{g}_{z\sigma}(\w)}{\partial \ln D}\! &=& \!
-2 \tilde{g}_\perp\!\left(B/2\right)^2 
\, \Theta(D-|\w+\sigma B|)
 \label{jRG} \\
\frac{\partial \tilde{g}_{\perp}(\w)}{\partial \ln D}\! &=&\!
-\!\!\! \sum_{\sigma=-1,1} \tilde{g}_\perp\!\left(\frac{\sigma B}{2}\right)
\tilde{g}_{z\sigma}(0)
\, \Theta\!\left(D-\left|{\w+\frac{\sigma B}{2}}\right|\right) \nonumber
\end{eqnarray}
with $\tilde{g}_{\perp}(\w)=\tilde{g}_{\perp}(-\w)$,
$\tilde{g}_{z\uparrow}(\w)=\tilde{g}_{z\downarrow}(-\w)$ and the
initial conditions $\tilde g_{z\sigma}(\omega) = \tilde
g_{\perp}(\omega) = J N_F$ at the bare cut-off $D_0$. The Theta
function $\Theta(\w)$ describes that various virtual processes are not
possible if the running cut-off $D$ gets too small. Note that one
recovers the usual poor man's scaling equations\cite{Anderson70} if
one considers only the limit $\w\to0$ as it is usually done in RG
schemes.

The RG equations (\ref{jRG}) are easily solved and we obtain for $D\to 0$
\begin{widetext}

\begin{eqnarray}
\tilde{g}_{z\sigma}(\w)&=&\Theta[|\w+\sigma B|-B] 
\frac{1}{2 \ln[|\w+\sigma B|/\TK]}+
\Theta[B-|\w+\sigma B|]  \frac{1}{2 \ln[B/\TK]} 
\left|\frac{B}{\w+\sigma B}\right|^{1/\ln[B/\TK]} \label{gg}\\
\tilde{g}_{\perp}(\w)&=&\sum_{\sigma} \Theta[|\w+{\sigma} \frac{B}{2}|-B] 
\frac{1}{4 \ln[|\w+{\sigma}\frac{B}{2}|/\TK]}
+ \Theta[B-|\w+{\sigma} \frac{B}{2}|]
\left(  \frac{1}{2 \ln[B/\TK]} 
\left|\frac{B}{\w+\sigma \frac{B}{2}}\right|^{\frac{1}{2 \ln[B/\TK]}}
\hspace*{-2mm} - \frac{1}{4 \ln[B/\TK]} \right)\nonumber
\end{eqnarray}

\end{widetext}
where within our one-loop approach $T_K=D_0 e^{-1/(2 N_F J)}$. As
expected, the dimensionless renormalized couplings are functions of
$\w/T_K$ and $B/T_K$ only.

It is important to realize that the logarithmic
renormalizations resummed in Eq. \eqref{jRG} are finally cut off by the
relaxation rate $\Gamma$ of the spin.  As the relevant processes involve at 
least one spin-flip, we identify $\Gamma$
with the transverse spin relaxation rate $1/T_2$, which is given in
terms of the renormalized couplings as\cite{RoschRG}
 \begin{multline}
 \Gamma= \frac{\pi}{4} \sum_{
  \gamma=-1,1}
 \int \!\! d\w \Bigl[
 \tilde{g}_{z\gamma}(\w)^2
  f_{\w}(1-f_{\w})\\
 +\tilde{g}_\perp(\w-\gamma B/2)^2
            f_{\w}(1-f_{\w- \gamma B})
 \Bigr].\label{goldenG}
 \end{multline}
 where $f_{\w}$ is the Fermi function.  The rate $\Gamma$ cuts off the
 power-law singularities at $\w=\pm B$ (or $\w=\pm B/2$) and we
 implement this by replacing $B/|\w+\sigma B|$ by
 $\sqrt{\Gamma^2+B^2}/\sqrt{\Gamma^2+(\w+\sigma B)^2}$ in (\ref{gg}).
 The errors introduced by this heuristic replacement are
 small\cite{Unpubl} in $1/\ln[\max(B,\w)/T_K]$.  For $B\gg T_K$, we
 obtain $\Gamma\approx \pi B /(16 \ln^2[B/T_K])\gg T_K$ and therefore
 $\tilde{g}_{z\sigma}(\w) \ll 1$ and $\tilde{g}_{\perp}(\w)\ll 1$.
 Accordingly, the perturbative RG is valid in this
 regime\cite{Rosch01} for {\em all} frequencies (while for $B\lesssim
 T_K$ it can be used only for $\w \gg T_K$).  The smallness of the
 renormalized couplings for either large $B$ or large $\w$ is the main
 reason why our calculation is controlled by $1/\ln[\max(\w,B])/T_K]$
 as e.g. higher-order corrections to (\ref{jRG}) remain small.

\begin{figure}
\includegraphics[width= 0.99\linewidth,clip=]{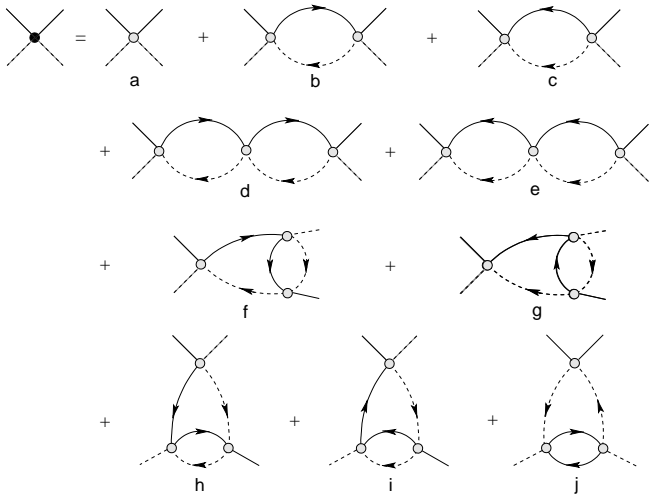}
\caption{\label{diagrams} Diagrams
  contributing to the vertex renormalization up to order $g^3$. Solid
  line: conduction electrons, dashed line: pseudo fermions representing
  the spin. See [\onlinecite{RoschRG}] for the corresponding Feynman
  rules. The (non-parquet) diagram (j) does not contribute to order
  $g^3 \ln^2 D_0$.}
\end{figure}

To check the assumptions underlying our formulation of the
perturbative RG for $\w$-dependent vertices, we have calculated the
vertices in next-to-leading logarithmic order in bare perturbation
theory by evaluating the diagrams shown in Fig.~\ref{diagrams}. Such a
test seems to be useful as RG schemes for $\w$-dependent vertices are
not used very often.  Neglecting terms of order $g^3\ln D_0$, we
obtain
\begin{widetext}
 \begin{eqnarray}
 \tilde{g}_{\perp}(\w)&\approx& g+ g^2 
\left(\ln\frac{D_0}{|\w\!-\!\frac{B}{2}|}+
\ln\frac{D_0}{|\w\!+\!\frac{B}{2}|}\right)+\frac{g^3}{2} \left(
\ln^2\!\frac{D_0}{|\w\!-\!\frac{B}{2}|}+6 
\ln \frac{D_0}{|\w\!-\!\frac{B}{2}|} \ln\frac{D_0}{|\w\!+\!\frac{B}{2}|} 
+\ln^2\!\frac{D_0}{|\w\!+\!\frac{B}{2}|}\right)\label{pert1}
\\
  \tilde{g}_{z\s}(\w)&\approx& g+ 2 g^2 \ln\frac{D_0}{|\w\!+\!\s B|}
+2 g^3  \left(     \ln \frac{D_0}{|\w|}\left[ \ln \frac{D_0}{|\w\!+\!\s B|} - 
          \ln \frac{D_0}{B}\right]
   + \ln \frac{D_0}{|\w\!+\!\s B|}\left[ \ln \frac{D_0}{|\w\!+\!\s B|} + 
          \ln \frac{D_0}{B}\right]  \right), \label{pert2}
\end{eqnarray}
\end{widetext}
where $g=N_F J$.  The perturbative RG is expected to resum {\em all} leading
logarithmic terms.  Indeed, if we expand (\ref{gg}) carefully in $g$
using that $T_K=D_0 e^{-1/(2 g)}$, we obtain the same logarithmic
divergences in all relevant limits, $\w \to \pm B/2$ or $\w \to \pm
B$, $\w \to \pm \infty$, and $\w \to 0$ [the singularity at $\w=0$ in
(\ref{pert2}) cancels]. One assumption which has not been checked in
the perturbative calculation shown above is that the logarithms are
cut off by the spin-relaxation rate. To analyze this effect one would
have to include self-energy insertions and corresponding vertex
corrections as will be discussed in a future publication\cite{Unpubl}.

\begin{figure}
\includegraphics[width= 0.95\linewidth,clip=]{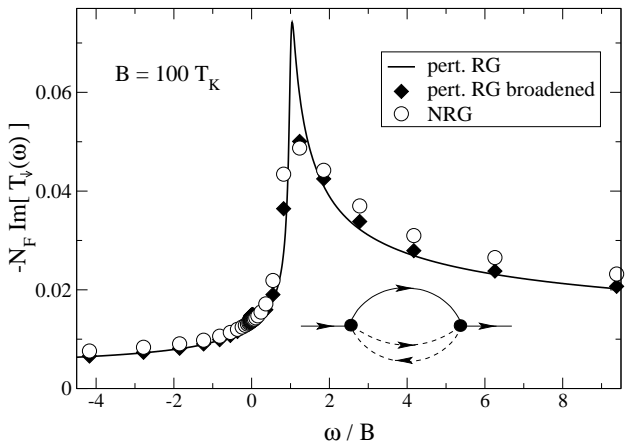}
\caption{\label{spectra100} $T$-Matrix 
  $-N_F \IM[T_{\downarrow}(\w)]$ (or equivalently the spectral
  function $-N_F V^2 \IM G^f_\downarrow(\w)$ of an Anderson impurity)
  as a function of $\w/B$ for $B=100\,T_K$. Solid line: perturbative
  RG, open circles: NRG.  The NRG is not able to resolve sharp peaks
  at high frequencies. For a direct comparison, we have therefore
  convoluted (filled diamonds) the result of the perturbative RG with
  the resolution function used in the NRG code. Inset: Diagram used to
  calculate the $T$-matrix (\ref{eom}) in leading order using renormalized
  vertices.}
\end{figure}

To leading order in $1/\ln[\max(\w,B)/T_K]$ the spectral function can
be calculated by replacing the bare coupling constants in the lowest
order expression (inset of Fig.~\ref{spectra100}) with the
renormalized vertex (\ref{vertex}) and one obtains\cite{keldysh}
\begin{multline}
N_F \text{Im}[ T_{{\sigma}}(\w)]=
-\frac{\pi}{4} \Bigl(  
\tilde{g}_{z{\sigma}}(\w)^2   \\+2
\tilde{g}_{\perp}(\w+{\sigma}\frac{B}{2})^2  
 \left[ 1+\sigma M ( 2 f_{\Gamma}(\w+{\sigma}B)-1)
\right]\Bigr)
\label{Tmatrix}
\end{multline}
where $M=2 \langle S_z \rangle$ is the magnetization of the impurity
normalized to $1$.  The Fermi function $f_\Gamma(\w\pm B)$ has to be
broadened by $\Gamma$, $f_\Gamma(\w)=1/2-\arctan[\w/\Gamma]/\pi$.
Dia\-gram\-ma\-ti\-cally this arises as the corresponding term in $T(\w)$
contains a convolution with the spin-flip susceptibility
$\chi^{+-}(\w)$ (see inset of Fig.~\ref{spectra100}) 
and we used the fact that the $\w$-dependence of the vertex
$\tilde{g}_\perp(\w)$ is sufficiently weak.

For $B\gg T_K$ one can replace $M$ in (\ref{Tmatrix}) by $1$ in
leading order of $1/\ln[B/T_K]$. However, to obtain the correct
large-$\w$ behavior for arbitrary $B$ it is essential to use the exact
local magnetization. For $B\sim T_K$ it is not possible to calculate
$M$ within our RG scheme, and we therefore use the magnetization
obtained from the Bethe Ansatz (or the practically identical values
from the NRG).

Friedel's sum rule\cite{friedel} relates the spectral function at
$\w=0$ exactly to the local magnetization $N_F \text{Im}[
T_{{\sigma}}(\w=0)]=-\frac{1}{\pi} \sin^2 \delta_\sigma$ where
$\frac{\delta_\sigma}{\pi}=\frac{1+\sigma M}{2}$. As a further check
of our approach, we compare this result to our perturbative RG. From
Eqs.~(\ref{gg}) and (\ref{Tmatrix}), we obtain asymptotically $N_F
\text{Im}[ T_{{\sigma}}(\w=0)]\approx -\frac{\pi}{16}
\frac{1}{\ln^2[B/T_K]}$ for $B\gg T_K$ which is
consistent with the asymptotic Bethe Ansatz\cite{natan}
result $M=1-1/(2 \ln[B/T_K])-\dots$

A typical spectral function for a spin-down electron is shown in
Fig.~\ref{spectra100}; the corresponding spin-up spectral function can
be obtained from $\IM T_{\uparrow}(\w)=\IM T_\downarrow(-\w)$.  For
large frequencies the spectral function 
decays as 
\begin{multline}
N_F \IM[T_\sigma(\w)] \approx 
-\pi \frac{3\mp 2 \sigma M}{16 \ln^2[\w/T_K]} \quad \text{for } \w \to \pm
\infty. \label{asym}
\end{multline}
A similar result, discussed below, has been obtained by Logan and
Dickens\cite{Logan,Logan2}.  

\begin{figure}
\includegraphics[width= 0.95\linewidth,clip=]{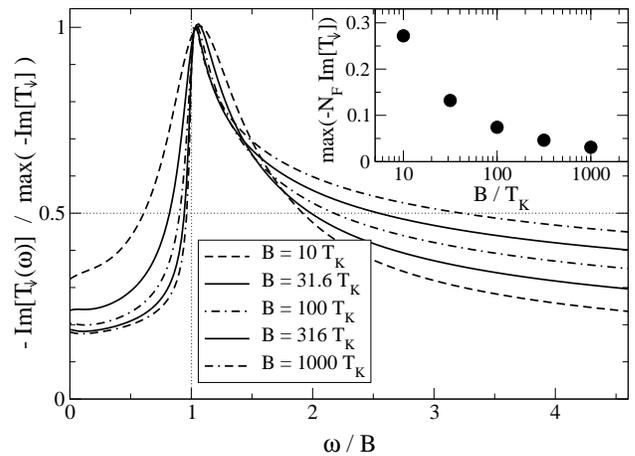}
\caption{ Line shape of the spectral function
  as a function of $\w/B$ calculated within perturbative RG. All
  curves are rescaled by their maximum. For increasing $B$ the peaks
  get more and more asymmetric.  Note that a small systematic shift of
  the peak position towards lower frequencies for smaller $B$ is not
  included (see text). Inset: peak height as a function of $B/T_K$.\label{shape}}
\end{figure}

For $\w\sim B$ the spectral function is characterized by a pronounced
and highly asymmetric peak. The width of the left flank is determined
by $\Gamma$ and is therefore of order $B/\ln^2[B/T_K]$. As $\Gamma/B\sim
1/\ln^2[B/T_K]$ decreases for increasing $B$, the left flank sharpens
with increasing $B$ (see Fig.~\ref{shape}). As the width of the right
flank increases for large $B$ the peak gets more and more asymmetric.
Formally, the line shape of the right flank is characterized by a
power-law for $\Gamma\ll \w-B \ll B$, see Eq.~(\ref{gg}).  However,
the corresponding exponents $\alpha$ of order $1/\ln[B/T_K]$ are so
small that it is not even in principle possible to extract the
power-law behavior\cite{extract} as $\alpha \ln[B/\Gamma]\sim
\ln[\ln(B/T_K)]/\ln(B/T_K) \ll 1$.

Within our perturbative RG, the peak in the spectral function is
always positioned at $\w=B$, see Fig.~\ref{shape}. This is the correct
$B\to \infty$ result. A subleading shift towards lower values is not
included in our scheme which neglects renormalization of the real part
of the self-energy\cite{Salmhofer01}. An estimate\cite{Unpubl}
suggests that this effect is of order $B/\ln[B/T_K]$ consistent with
results from Moore and Wen\cite{MooreWen} discussed below.  For $B\ll
T_K$ it has been shown\cite{Logan,Hewson01,costi.02} that the peak in
the spectral function is positioned at $2 B/3$. A splitting in the
total spectral function $\IM T_\uparrow(\w)+\IM T_\downarrow(\w)$ can
be observed\cite{costi.00} for $B\gtrsim 0.5\, T^*_K$, where 
$T^*_K\approx 2.8 \,T_K$ is the half width at half maximum (HWHM) of the Kondo 
resonance at $T=0$.

\section{NRG}
The results of the perturbative RG approach to the equilibrium
$T$-matrix $T(\omega)$ can be compared to numerical renormalization
group (NRG) results for the same quantity \cite{costi.00}.  The NRG
procedure \cite{wilson.75+kww.80} consists of the following steps: (i)
a logarithmic mesh of k-points $k_{n}=\Lambda^{-n}$ is introduced
about the Fermi wavevector $k_{F}=0$, and (ii) a unitary
transformation of the $c_{k\sigma}$ is performed such that
$f_{0\sigma}=\sum_{k}c_{k\sigma}$ is the first operator in a new
basis, $f_{n\sigma},n=0,1,\dots$, which tridiagonalizes
$H_{c}=\sum_{k\sigma}\epsilon_{k\sigma}c_{k\sigma}^{\dagger}c_{k\sigma}$
in $k$-space, i.e.  $H_{c}\rightarrow
\sum_{\sigma}\sum_{n=0}^{\infty}\Lambda^{-n/2}
(f_{n+1\sigma}^{\dagger}f_{n\sigma}+ h.c.).$ The Hamiltonian
(\ref{kondo}) with the above discretized form of the kinetic energy is
now iteratively diagonalized by defining a sequence of finite size
Hamiltonians $H_{N} = \sum_{\sigma,n=0}^{N-1}\Lambda^{-n/2}
(f_{n+1\sigma}^{\dagger}f_{n\sigma}+ h.c.) + J \sum_{\s,\s'}\vec{S}
\cdot f^{\dagger}_{0,\s'} {\boldsymbol \tau}_{\s'\s}f_{0,\s}- B S_{z}
$ for $N\ge 0$.  For each $N$, this yields the excitation energies
$E_{\lambda}^{N}$ and many-body eigenstates $|\lambda\rangle_{N}$ of
$H_N$ at a corresponding set of energy scales $\omega_{N}$ defined by
the smallest scale in $H_N$, $\omega_{N}=\Lambda^{-\frac{N-1}{2}}$.
Since the number of states grows as $4^N$, for $N>6$ only the lowest
500 or so states are retained for $H_N$. This limits the width of the
spectrum of $H_N$ to $0\leq \omega \leq K(\Lambda)\omega_N$, where
$K(\Lambda)\approx 5$ for the value $\Lambda=1.5$ used in this paper.
We are interested in $\IM T(\omega)= \IM \langle\langle O; O^\dagger
\rangle\rangle_{\rho}$, where $O= \vec{S} c^\dagger_\alpha
{\boldsymbol \tau}_{\alpha \s}$ is defined in Eq.\ (\ref{eom}) and the
subscript $\rho$ indicates the density matrix used to evaluate the
thermodynamic averages (see below).  The matrix elements $\langle
m|O|n\rangle_{N}$, which are required for $T(\omega)$, are also
calculated iteratively. $\IM T(\omega)$ is then constructed at a
characteristic set of frequencies $\omega = \Omega_{N}$ from $H_{N}$
for each $N=0,1,\dots$ via its Lehmann representation. This
necessarily yields $\IM T(\omega)$ in the form of a set of weighted
delta functions $\delta(\Omega_N \pm E_{\lambda}^{N})$.  Continuous
spectra are obtained by replacing the delta functions by Gaussians of
width $\eta_N$: $ \delta(\Omega_N \pm E_\lambda^{N})\rightarrow
\frac{1}{\eta_N \sqrt{\pi}} \exp(-(\Omega_N \pm
E_\lambda^{N})^2/\eta_N^2) $. As noted above, the spectrum of $H_N$ is
known in a finite window of size $\approx 5\omega_N$ for
$\Lambda=1.5$.  We therefore choose $\Omega_N=1.8\omega_N$, not too
low since the lower part of the spectrum is refined in later
iterations and not too close to the upper bound of the spectrum which
could be subject to truncation errors. We also
set $\eta_N=0.39\, \Omega_N$ 
of order the level structure of $H_N$.  The main interest in this
paper is to compare the NRG results for $\IM T(\omega,B,T=0)$ with the
perturbative RG calculations in the limit $B\gg T_K$ where the latter
are expected to be accurate. In this limit, the spectral features in
$\IM T(\omega,B,T=0)$ lie at high frequencies $\omega\approx B$.  A
description of this finite-field, high-frequency part of the spectrum
within NRG requires the use of a reduced density 
matrix\cite{hofstetter.00,costi.02} calculated
from the true ground state, i.e.  $H_{N\to \infty}$, in place of the
grand-canonical density matrix obtained from $H_N$. The necessity to use
ground-state properties to calculate high-frequency features is for
example evident from Eq.~(\ref{asym}): the magnetization determines
the large-$\w$ tails.  Nevertheless, the two approaches have been
shown to give almost identical results at $B=0$ and to be in good
agreement\cite{costi.02} for fields $B\lesssim 10 T_K$.

 \begin{figure}
 \includegraphics[width= 0.95 \linewidth,clip=]{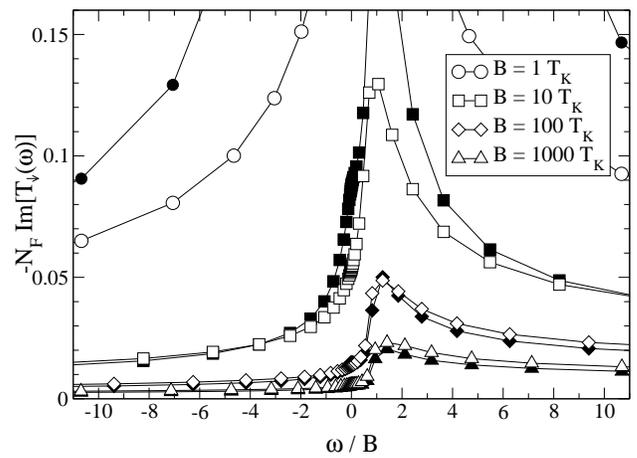}
 \caption{\label{figSpectraB} Comparison of NRG (open symbols) 
   and perturbative RG (filled symbols) broadened by the NRG-resolution
   function  (see caption of Fig.~\ref{spectra100}) for large magnetic
   fields, $B=1,10,100,1000 \, T_K$.}
 \end{figure}

\section{Comparisons}
\begin{figure}
\begin{center}
\includegraphics[width=0.95 \linewidth,clip=]{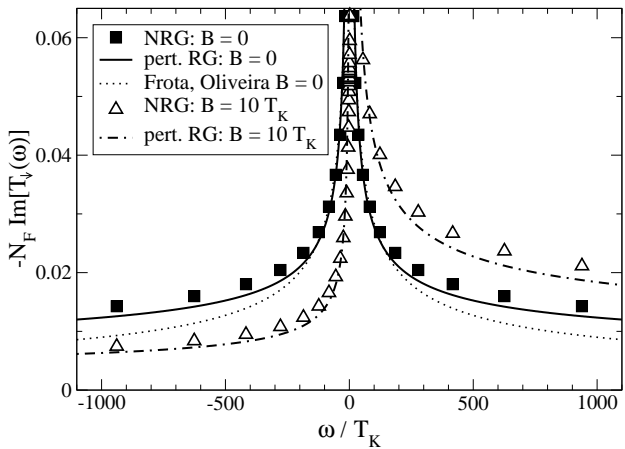}
\end{center}
\caption{\label{largeWfig} Behavior of the spectral function 
$-N_F \IM[T_\downarrow(\w)]=-N_F V^2 \IM G^f_\downarrow(\w)$
  for large $\w$ and $B=0,10 \, T_K$. 
For $B,T_K\ll \w$ the spectral
  function decays as $1/\ln^2[\w/T_K]$ with prefactors depending on
  the magnetization [see Eq.~(\ref{Tmatrix})]. Symbols: NRG, solid
  and dot-dashed line: perturbative RG. For very large $\w>200\, T_K$
  one can see that an interpolation formula suggested in
  Ref.~\onlinecite{Frota} (dotted line) decaying as $1/\sqrt{\w}$ is
  not consistent\cite{Logan2} with the NRG data.  }
\end{figure}

How well does the perturbative RG reproduce the spectral functions
calculated within NRG?  Fig.~\ref{spectra100} shows excellent
agreement\cite{TkNRG} for $B/T_K=100$ both for $\w \to 0$ and $\w \to
\pm \infty$. However, large deviations arise for $\w \approx B$ where
the peak within the perturbative RG is much sharper and higher.  This
disagreement arises because the logarithmic discretization in
energy used within NRG cannot resolve sharp features at high
frequencies.  More precisely, for a feature at a given frequency
$\Omega_0$ with intrinsic width $\Delta_0$ it will fail to resolve
it\cite{costi.99} if $\Omega_0 > \Delta_0$ since the broadening used
at frequency $\Omega_0$ is necessarily of order $\Omega_0$ ($0.39\,
\Omega_0$ in our calculations\cite{broadening}).  
This is the case here, since the
spectral feature at $\Omega_0=B$ has a width on its left flank given
by $\Delta_0\sim B/\ln^2[B/T_K]$. Consequently, the width of this
feature will not be captured by NRG for $B\gg T_K$.  To allow a direct
comparison of NRG and perturbative RG, we have broadened the latter at
the energy $\w$ by convolution with the resolution function used in
the NRG code, i.e. by a Gaussian of width $0.39 \, \w$.  After this
broadening, the agreement of perturbative RG and NRG is excellent for
all frequencies as long as $B\gg T_K$, see Figs.~\ref{spectra100} and
\ref{figSpectraB}. For smaller $B$, the spectral function cannot be
calculated reliably for small $\w$ within our perturbative RG and
indeed strong deviations can be seen in Fig.~\ref{figSpectraB} for
$B\lesssim 10\, T_K$.  Fig.~\ref{largeWfig} shows that the large $\w$
behavior is nevertheless well reproduced which firmly establishes
Eq.~(\ref{asym}). Note that we replaced $M$ in (\ref{Tmatrix}) by the
true ground-state magnetization for this comparison as discussed
above.

\begin{figure}
\begin{center}
\includegraphics[width=0.95 \linewidth,clip=]{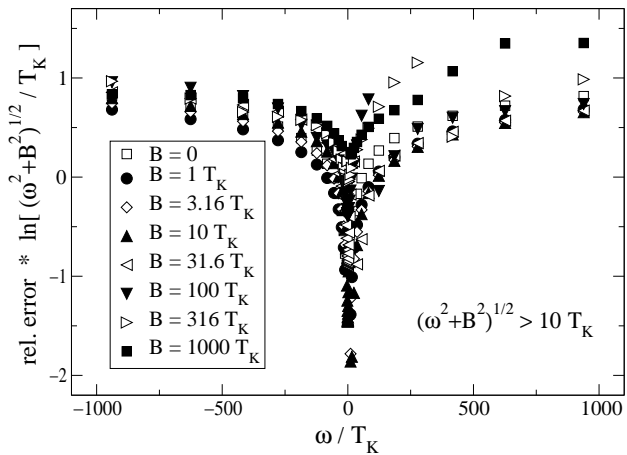}
\end{center}
\caption{\label{error} The perturbative RG is a controlled expansion 
  in the small parameter $1/\ln[\max[\w,B]/T_K]$. This is shown by
  plotting the relative error of the perturbative RG compared to NRG,
  $\frac{\text{NRG-pert.RG}}{\text{NRG}}$, multiplied by
  $\ln[\sqrt{\w^2+B^2}/T_K]$ for various frequencies and magnetic
  fields. The perturbative RG results have been broadened by the NRG
  resolution (see Fig.~\ref{spectra100} and text). Only data points
  with $\sqrt{\w^2+B^2}>10 \, T_K$ are shown.  }
\end{figure}

We have argued that the small parameter controlling the validity of
our perturbative RG is $1/\ln[\max(B,\w)/T_K]$. To check this claim we
plot in Fig.~\ref{error} the relative error of the perturbative RG
(compared to NRG) multiplied by $\ln[\sqrt{\w^2+B^2}/T_K]$ for a wide
range of $\w$ and $B$. As expected, one obtains a number of order unity!
Fig.~\ref{error} suggests that the next-order correction is suppressed
by a factor smaller than $1.5/\ln[\max(B,\w)/T_K]$. 
From the large-$B$ expansion of quantities like the magnetization
calculated by  Bethe Ansatz\cite{natan,Hewson93} one
expects that the correction rather has the form
$\ln[\ln[\max(B,\w)/T_K]]/\ln[\max(B,\w)/T_K]$ but our range of parameters
is too small to extract a $\ln[\ln[\dots]]$ prefactor.

In the remainder of this section, we briefly compare our results to
other approaches.  Many years ago, Frota and Oliveira\cite{Frota}
described the spectral function for $B=0$ with the simple heuristic
form $N_F T(\w)= -1/\RE[\pi (1-i \w/\Gamma_K)^{1/2}]$ where
$\Gamma_K\approx 0.4 \, T_K$ (in our units) was fitted to the width of
the Kondo peak.  This form was chosen to interpolate between the exact
$\w=0$ result and an asymptotic $1/\sqrt{\w}$ behavior which was
believed to arise as a consequence of an X-ray singularity in the
presence of a phase-shift $\pi/2$.  The fit to the Frota and Oliveira
formula works surprisingly well for frequencies up to $200 \, T_K$
(see Fig.~\ref{largeWfig}).  From our point of view, this remarkable
agreement is, however, accidental in the following sense: For $T_K\ll
\w \ll D$ the spectral function decays as $1/\ln^2[\w/T_K]$ and not as
$1/\sqrt{\w}$ as has previously been pointed out by Dickens and 
Logan\cite{Logan2}.  Does a $1/\sqrt{\w}$ law hold at intermediate
frequencies, e.g. for $T_K \ll \w \ll T^*$?  This would require the
existence of a new scale $T^*$ parametrically larger than $T_K$ in
contradiction not only to the perturbative RG but also to the exact
Bethe Ansatz solution\cite{natan}.  Nevertheless, the Frota-Oliveira
formula is a very good heuristic description in a large regime.  For
$20 \, T_K <\w<200 \, T_K$ it is almost indistinguishable from our
perturbative RG result for $B=0$ (and fits the exact
result much better for $\w<20 \, T_K$ where our perturbative scheme
breaks down).

Recently, Logan  and Dickens\cite{Logan,Logan2}
 calculated the spectral functions in a
magnetic field within their ``local moment approach''  (LMA), which is
based on a combination of diagrammatic perturbation theory and
mean-field  theory. Interestingly, Logan and Dickens obtain the correct
$\omega \to \infty$ and $\omega \to 0$ asymptotics. However the peak in the
spectral function close to $\omega =B$ appears to be much broader in their
approach and they find that the peak is shifted towards larger frequencies
of order $ B \ln(B/T_{K})$. 
We think that both features are an artifact of their
heuristic scheme to introduce broadening.

Moore and Wen \cite{MooreWen} calculated the density of states of the
spinons obtained from the Bethe Ansatz solution\cite{natan}. They
suggested that their result may serve as an approximation to the
spectral function.  However, the spinon density of states lacks the
logarithmic tails characterizing the large $\w$ behavior of the
spectral function and their peaks close to $\w=B$ do not show the
correct asymmetric line shape for $B\gg T_K$. Nevertheless, it is
interesting to note that the width of the spinon density of states is
of order $\sim B/\ln^2[B/T_K]$ like the left flank of our result.  In
their result, the position of the peak in the spectral function is
shifted by $-B/(2\ln[B/T_K])$ away from $\w=B$ for $B\gg T_K$.  As
discussed above, such a shift is beyond the precision of our present
calculation but an interesting open question is whether the
peak-position in the spinon density of states coincides with that in
the electron spectral function.

\section{Conclusion}

In this paper we have addressed three problems in particular. 

(i) We have calculated analytically the behavior of the universal
spectral function in the Kondo model for either large frequencies or
large magnetic field and arbitrary frequencies. We hope that our
results establishes the existence and detailed shape of the highly
asymmetric peak at frequencies of the order of the Zeeman splitting
and the asymptotic behavior at high frequencies. Hopefully, the
spectral function in a magnetic field can be studied in some detail in
future experiments e.g. by tunneling into magnetic impurities.

(ii) In
a detailed comparison with perturbation theory and NRG, we have argued that
our recently developed perturbative RG scheme\cite{RoschRG} allows
controlled calculations in the small parameter
$1/\ln[\max(\w,B)/T_K]$. We expect that a similar statement holds out of
equilibrium for large bias voltages $V\gg T_K$ where a comparison
with numerically exact results is not yet possible. 

(iii) We
have tried to assess the ability of the NRG to resolve structures at high
frequencies. As is obvious from the formulation of NRG, structures
which are much sharper than their typical frequency are not correctly
reproduced by NRG.  However, our results suggest that the error
induced by the logarithmic discretization mainly results in a simple
broadening. 
The spectral function for large $B$ can serve as a test for future
improvements of the NRG algorithms adapted to describe high-$\w$
features more precisely, which may be useful, for example, in
implementations of dynamical mean field theory using
NRG\cite{bullaDMFT}.

We thank J. Kroha, D.E. Logan, J. Martinek and M.~Vojta for helpful
discussions and especially L.~Glazman who motivated us to check the
validity of our RG approach in more detail. This work was supported in
part by the CFN, the Emmy Noether program (A.R.) of the DFG and the
SFB 195 of the DFG (TAC). We also thank the {\O}rsted Laboratory for
hospitality during parts of this work (J.P.).


\begin{thebibliography}{0}
  

\bibitem{STM} 
J. Li, W.-D. Schneider, R. Berndt, and B. Delley, Phys. Rev. Lett. 
{\bf 80}, 2893 (1998), 
V. Madhavan, W. Chen, T. Jamneala, M. F. Crommie, and N. S. Wingreen,
Science {\bf 280}, 567 (1998); 
H. C. Manoharan, C. P. Lutz, and D. W.  Eigler, Nature {\bf 403}, 512 (2000).
N. Knorr, M. A. Schneider, L. Diekh\"oner, P. Wahl, and K. Kern, 
Phys. Rev. Lett. {\bf 88}, 096804 (2002). 

\bibitem{Goldhaber98}D. Goldhaber-Gordon, H. Shtrikman, 
D. Mahalu, D. Abusch-Magder, U. Meirav, and M.A. Kastner,
Nature {\bf 391}, 156 (1998);
S. M. Cronenwett, T. H. Oosterkamp and L. P. Kouwenhoven,
Science {\bf 281}, 540 (1998);
W. G. van der Wiel, van der Wiel, S. De Franceschi, T. Fujisawa,
J.M. Elzerman, S. Tarucha, and L.P. Kouwenhoven,
Science {\bf 289}, 2105 (2000);
J. Nyg{\aa}rd, D. H. Cobden and P. E. Lindelof,
Nature {\bf 408}, 342 (2000).

\bibitem{Franceschi}
S. De Franceschi, R. Hanson, W. G. van der Wiel, J. M. Elzerman, 
J. J. Wijpkema, T. Fujisawa, S. Tarucha, and L. P. Kouwenhoven 
Phys. Rev. Lett. {\bf 89}, 156801 (2002). 


\bibitem{Hewson93}A. C. Hewson, {\it The Kondo Problem to Heavy Fermions},
Cambridge University Press (1993).

\bibitem{dephasing}
J. Kroha, Adv. Solid State Phys. {\bf 40}, 216 (2000); A.~Ka\-min\-ski, 
L.~I. Glazman, Phys. Rev. Lett. {\bf 86}, 2400 (2001);   
G. G\"oppert and H. Grabert, Phys. Rev. B {\bf 64}, 033301 (2001);
J. Kroha and A. Zawadowski, Phys. Rev. Lett. {\bf 88}, 176803 (2002).

\bibitem{DMFT}  W. Metzner and D. Vollhardt, Phys. Rev. Lett. {\bf 62}, 
324 (1989); A. Georges, G. Kotliar, W. Krauth, and
M.~J. Rozenberg, Rev.  Mod. Phys. {\bf 68}, 13 (1996).

\bibitem{Meir}Y. Meir, N.~S. Wingreen, and P.~A. Lee, 
Phys. Rev. Lett. {\bf 70}, 2601 (1993).

\bibitem{Sakai} 
O. Sakai, S. Suzuki, W. Izumida, and A. Oguri,  
J. Phys. Soc. Jap. {\bf 68}, 1640 (1999).

\bibitem{Takagi} O. Takagi and T. Saso, 
J. Phys. Soc. Jap. {\bf 68}, 1997 (1999).

\bibitem{costi.00} T. A. Costi, Phys. Rev. Lett. {\bf 85}, 1504 (2000).



\bibitem{hofstetter.00} W. Hofstetter, Phys. Rev. Lett. {\bf 85}, 1508 (2000).


\bibitem{MooreWen} 
J. E. Moore and X. G. Wen, Phys. Rev. Lett. {\bf 85}, 1722 (2000).

\bibitem{Logan} D. E. Logan and N. L. Dickens, Europhys. Lett. {\bf
    54}, 227 (2001), D. E. Logan and N. L. Dickens, J. Phys.: Cond.
  Mat. {\bf 13}, 9713 (2001).

 
\bibitem{Hewson01}
A. C. Hewson, J. Phys.: Cond. Mat. {\bf 13}, 10011 (2001).

\bibitem{Martinek}  
  J. Martinek, Y. Utsumi, H. Imamura, J. Barnas, S. Maekawa, J. K\"onig, 
and G. Sch\"on, preprint cond-mat/0210006; 
J. Martinek, M. Sindel, L. Borda, J. Barna\'s, J. K\"onig, G. Sch\"on, 
J. von Delft, preprint cond-mat/0304385.

\bibitem{wilson.75+kww.80} K. G. Wilson,  Rev. Mod. Phys. {\bf 47}, 773
(1975); H. R. Krishna-murthy, J. W. Wilkins and K. G. Wilson,  
Phys. Rev. {\bf B21}, 1003 (1980).


\bibitem{Frota} H. O. Frota and 
L. N.  Oliveira, Phys. Rev. B {\bf 33}, R7871 (1986).

\bibitem{sakai.89} O. Sakai, Y. Shimizu and T. Kasuya, J. Phys. Soc.
Jap. {\bf 58} 3666 (1989)



\bibitem{costi.94} T. A. Costi, A. C. Hewson and V. Zlati\'c,
J. Phys. Cond. Matt. {\bf 6} 2519 (1994).



\bibitem{RoschRG}A. Rosch, J. Paaske, J. Kroha and P. W\"{o}lfle,
Phys. Rev. Lett. {\bf 90}, 076804 (2003), 




\bibitem{magneticField} Adding a magnetic field also to the conduction
  electrons will not change the results in the scaling limit as the
  density of states of the up and down electrons at the Fermi energy
  remains independent of $B$.
 

 
\bibitem{Unpubl}A. Rosch, J. Paaske, J. Kroha, P. W\"olfle, unpublished.
  

\bibitem{pseudo} The vertices are defined using a pseudo spin representation of the spin, see
Ref.~\onlinecite{RoschRG}.

\bibitem{Anderson70}P. W. Anderson,
J. Phys. C {\bf 3}, 2436 (1970).

\bibitem{Rosch01}A. Rosch, J. Kroha and P. W\"{o}lfle,
Phys. Rev. Lett. {\bf 87}, 156802 (2001).


\bibitem{keldysh} While we are considering only equilibrium
  quantities, the proper replacements of bare and renormalized
  quantities is most easily done in the Keldysh formulation where it
  is easy to identify correctly the impurity magnetization in
  (\ref{Tmatrix}). In an equilibrium technique it would be far from
  obvious that $\tanh(B/2 T)$ has to be replaced by $M$.


\bibitem{friedel} D.~C. Langreth, Phys. Rev. {\bf 150}, 516 (1966);
P. Nozi\`eres, J. Low Temp. Phys. {\bf 17}, 31 (1974).

\bibitem{natan} N. Andrei, Phys. Rev. Lett. {\bf 45}, 379 (1980);
N. Andrei, Phys. Lett. {\bf 87A}, 299 (1982).

\bibitem{Logan2} N. L. Dickens and D. E. Logan, J. Phys.: Cond. Mat.
 {\bf 13}, 4505 (2001).

\bibitem{extract} To extract a power law $x^{\alpha}=e^{\alpha \ln x}$
  the variations of the  
argument of the $e$ function should be large compared to $1$.

\bibitem{Salmhofer01} 
M. Salmhofer and C. Honerkamp, Prog. Theo. Physics {\bf 105}, 1 (2001).


\bibitem{costi.02} T. A. Costi, proceedings of NATO ARW on ``Concepts in
Electron Correlations'', Hvar, Croatia, edited by  
A. C. Hewson and V. Zlati\'c (Plenum Press, New York, 2003).

  
\bibitem{TkNRG} For our NRG results we use the asymptotically exact
  two-loop definition $T_K=D_0 \sqrt{2 N_F J} e^{-1/(2 N_F J)}$ of the
  Kondo temperature to be identified with $T_K=D_0 e^{-1/(2 N_F J)}$
  within our one-loop perturbative RG.


\bibitem{costi.99} T. A. Costi, in {\em Density Matrix 
Renormalization}, edited by I. Peschel, X. Wang, M. Kaulke and 
K. Hallberg (Springer, Berlin, Germany 1999.

\bibitem{broadening} As long as a logarithmic discretization of
  energies is used within NRG, the use of a different broadening
  scheme will not improve the agreement significantly. However, in
  some specific cases, an improvement in resolving finite energy peaks
  within NRG can be obtained by making use of the self-energy as
  described in: R. Bulla, A. C. Hewson and Th. Pruschke, J. Phys.
  Cond. Matt. {\bf 10}, 8365 (1998).


\bibitem{bullaDMFT} R. Bulla, Phys. Rev. Lett. 83, 136 (1999); 
R. Bulla, T.~A.~Costi and D. Vollhardt, Phys. Rev. 
B {\bf 64}, 045103 (2001); 
R. Zitzler, Th. Pruschke and R. Bulla, 
Eur. Phys. J. B {\bf 27}, 473 (2002); 
T. A. Costi and N. Manini, 
J. Low Temp. Phys. {\bf 126}, 835 (2002).



 




\end{thebibliography}
\end{document}